\documentclass[article]{article} 

\usepackage{tabularx}
\usepackage{lscape}   
\usepackage{longtable}
\usepackage{graphicx}
\usepackage{array}
\usepackage{booktabs} 
\usepackage{amsmath,amssymb}
\usepackage{pdflscape} 
\usepackage{multirow}
\usepackage{makecell}

\graphicspath{{figs/}}

\title{Dimples for skin-friction drag reduction: \\ 
status and perspectives}

\author{Federica Gattere, Alessandro Chiarini and Maurizio Quadrio \\
\small Politecnico di Milano, Dept. of Aerospace Science and Technologies\\
}
\date{}

\begin{document}

\maketitle

\abstract{Dimples are small concavities imprinted on a flat surface, known to affect heat transfer and also flow separation and aerodynamic drag on bluff bodies when acting as a standard roughness. Recently, dimples have been proposed as a roughness pattern that is capable to reduce the turbulent drag of a flat plate, by providing a reduction of skin friction that compensates the dimple-induced pressure drag, and leads to a global benefit. 
\\ 
The question whether dimples do actually work to reduce friction drag is still unsettled. In this paper, we provide a comprehensive review of the available information, touching upon the many parameters that characterize the problem. A number of reasons that contribute to explaining the contrasting literature information are discussed. We also provide guidelines for future studies, by highlighting key methodological steps required for a meaningful comparison between a flat and dimpled surface in view of drag reduction.
}

\section{Introduction}

Reducing the drag generated by a fluid set in relative motion to a solid body is at the same time a fundamental attempt to learn how to favorably interact with turbulence, and a technological challenge with immense potential in so many application fields. The interest for turbulent flow control is steadily increasing, owing to massive economic and environmental concerns. 

Skin-friction drag is perhaps the most essential manifestation of the dissipative nature of turbulence, and accounts for the total drag in the case of planar walls (as in a channel flow or a zero-incidence flat plate boundary layer). Several techniques are available to reduce friction drag below the level typical of a smooth solid wall; they can be categorized into active (requiring extra energy) and passive ones.
The former typically provide larger savings, but imply extra complexity and cost, so that the ideal technique for friction reduction remains a passive one, often embodied in surface patterns performing better than the planar flat geometry.

The most prominent example of such patterns is riblets \cite{garcia-jimenez-2011-a}. Introduced by NASA in the '80 of the past century, and intensely studied over the subsequent years for their potential in aeronautical applications, riblets consist of streamwise-aligned microgrooves, and have the proved ability to reduce friction drag. The riblets cross-section can be of several shapes, the triangular one being perhaps the most popular, but an essential feature is a very sharp tip. Although new developments \cite{quadrio-etal-2022,cacciatori-etal-2022} hint at a bright future for riblets in aeronautics and suggest lower cost/benefit ratios, riblets are currently still not deployed in commercial transport aircraft, owing to their limited savings \cite{mclean-georgefalvy-sullivan-1987, kurita-etal-2020} and to critical production and maintainance issues, descending from the microscopic size of riblets and from the requirement of preserving the sharpness of the tip. 

A possible alternative to riblets is emerging recently, easier to manufacture and lacking any sharp detail. The pattern to impress on the surface consists of small dimples. Dimples, i.e. small concavities imprinted on a surface, have been extensively studied in the past for their ability to enhance the heat transfer of a surface (see e.g. ref.\cite{leontiev-etal-2017} and references therein). The use of dimples on the surface of bluff bodies (e.g. a golf ball) is well known, and their ability to influence the turbulent boundary layer and the separation on the body is rather well understood \cite{choi-jeon-choi-2006}; the same concept is also being considered in sport-car racing \cite{allarton-etal-2020}. In this paper we concern ourselves with dimples applied to an otherwise flat surface: the goal is to reduce the turbulent skin-friction drag. We limit our review to passive dimples, although also active control by dimples has been proposed \cite{ge-fang-liu-2017}.

The ability of dimples to reduce drag is way less intuitive than that to increase heat exchange, and was considered first at the Kurchatov Institute of Athomic Energy \cite{kiknadze-krasnov-chushkin-1984} in Russia, where hemispherical dimples were placed on the surface of a heat exchanger and found to reduce the flow resistance as well. 
In subsequent studies by the same group, a drag reduction of about 15--20\% was mentioned \cite{alekseev-etal-1998}, the highest performance reported so far. 
In the last two decades, a handful of research groups devoted their efforts to understanding the drag reduction problem by dimples, attempting to come up with a recipe for the best shape and size.
Unfortunately, to date no consensus has been reached on the effectiveness of dimples in reducing friction drag, and on their working mechanism: some authors confirmed drag reduction, others did not.

Measuring -- in the lab, or with a numerical simulation -- the (very small, if any) drag reduction induced by dimples is by no means a trivial task. 
A reduction of friction drag would be unavoidably accompanied by an increase of pressure drag, with a net benefit possible only if the former overwhelms the latter. 
There are just so many design variables to be tested, as the geometry of the dimple itself, the size, the spatial layout and relative arrangement of the dimples on the surface need to be carefully considered. This is a daunting task as long as no theory, hypothesis of working mechanism or scaling argument is available to guide the search in such a vast parameter space. 
However, it is undeniable that dimples, once proved to work, would provide substantial advantages over riblets, thanks to their simplicity, ease of production, lack of sharp corners and easier maintenance.  

The goal of the present contribution is to provide the first comprehensive review of the published information available on dimples for skin-friction drag reduction.
Since the very fact that dimples can actually work is still subject to debate, we will complement the review with a discussion of important procedural aspects that in our view are essential, should one embark in a (numerical or laboratory) experiment to assess the potential for drag reduction. 
Such procedures (or, more precisely, their absence) are at the root of the large uncertainty and scatter of the available data, and have hindered so far the answer to such a simple question as: Do dimples actually work to reduce turbulent drag?

The present contribution is structured as follows. Section \S\ref{sec:overview} provides an overview of the experimental and numerical studies on the drag reduction properties of dimples. Section \S\ref{sec:parameters} describes the geometrical parameters defining the dimples, whereas \S\ref{sec:physics} reports the two main physical explanations for the working mechanism of drag-reducing dimples. In \S\ref{sec:analysis} we highlight the problem of properly measuring drag reduction, and guidelines and recommendations on how to properly compare results among different studies are provided. This review paper is closed by brief concluding remarks in \S\ref{sec:conclusions}. Appendix \ref{sec:appendix} contains details of the DNS simulations that have been carried out for the present study.  

In the next Subsection, the concept of dimples is introduced first, together with the notation that will be used later to indicate their geometrical parameters.

\subsection{Characterization of a dimpled surface}

\begin{figure}
\includegraphics[width=1\textwidth]{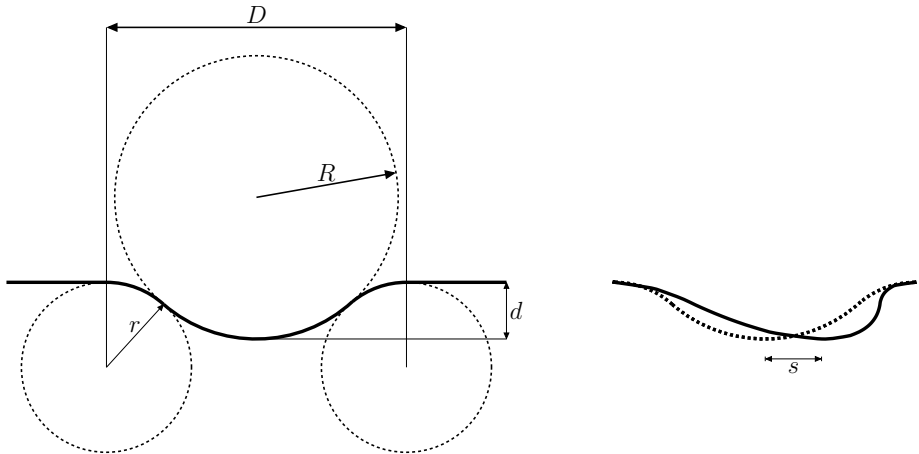}
\caption{Cross-section of the parametric dimple geometry introduced by ref.\cite{chen-chew-khoo-2012} (left) and streamwise shift of the deepest point (right).}
\label{fig:spherical_geom}
\end{figure}

A solid wall covered with dimples is described by several geometric parameters: the dimple shape, the relative spatial arrangement of the dimples and the coverage ratio (ratio between non-planar and total surface). Originally, dimples were conceived as spherical recesses, hence with a circular footprint on the wall. One particular class of circular dimples, introduced by Chen et al. \cite{chen-chew-khoo-2012}, has become quite popular thanks to its parametric nature and represents the starting point of our description. This design is the union of a spherical indentation and a torus, meeting tangentially in a regular way that avoids sharp edges. A cross-section of this dimple, which possesses axial symmetry, is drawn in figure \ref{fig:spherical_geom}. 

Four parameters define the geometry of this dimple: the diameter $D$ of the circular section at the wall, the depth $d$ of the spherical cap, the curvature radius $r$ at the edge and the curvature radius $R$ of the spherical cap. These four parameters are not independent, but linked by one analytical relation, so that only three degrees of freedom exist. In fact, geometry dictates that:
\begin{equation}
\frac{D}{2} = \sqrt{ d \left( 2R + 2 r -d \right)}.
\label{eq:geom_rel}
\end{equation}

Moreover, a handful of studies extended this baseline circular geometry, by introducing the additional parameter $s$, which describes the shift along the streamwise direction (either downstream for $s>$ or upstream for $s<0$) of the point of maximum depth, which in the baseline geometry lies exactly at the center of the dimple cavity. 

\begin{figure}
\centering
\includegraphics[width=\textwidth]{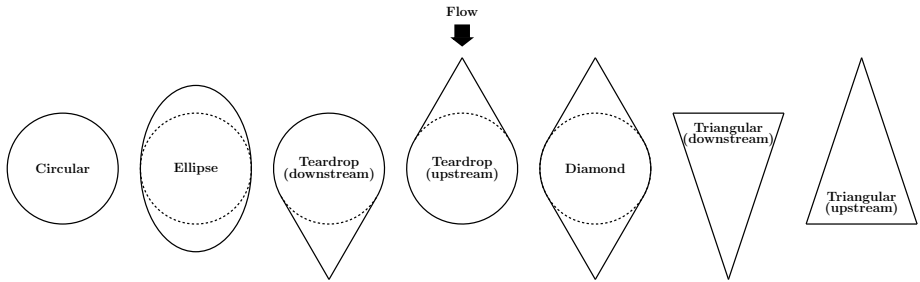}
\caption{Popular variants of the dimple shape.}
\label{fig:diff_geom}
\end{figure}

It is difficult to overemphasize the importance of a well-defined parametric geometry in the quest for the optimally performing dimple. Although the circular shape is by far the most popular, over the last years a number of alternative dimple shapes have been studied; sketches for the various shapes are drawn in figure \ref{fig:diff_geom}. Some of them derive from a deformation of the circular shape: e.g. the elliptical dimple is the result of a symmetrical stretch of the circular dimple along the streamwise direction. The teardrop dimple has two segments tangent to the circle, preserves the spanwise symmetry and exists in two variants depending on whether the triangle points upstream or downstream. The diamond dimple is the union of the two variants of teardrop and possesses two vertices. Only the triangular dimple differs substantially from the circular shape and --- as for the teardrop dimple --- can have the streamwise-aligned vertex pointing upstream or downstream.

When a single dimple is identically replicated to fully cover the planar surface, the relative spatial arrangement of the dimples is important in determining the overall influence on the flow. 
A regular spatial layout of a dimpled surface depends on the distance between two adjacent dimples in both the streamwise and spanwise directions. 
Another parameter that is related to the spatial arrangement of dimples is the coverage ratio, that can be defined as the percentage of recessed surface compared to the total surface of the wall. 
(The reader will notice an ambiguity, as at the denominator of the coverage ratio one could put either the surface area of the equivalent flat wall, or the wetted area of the dimpled surface. This ambiguity is often ignored, but it is commented upon e.g. in refs.\cite{prass-etal-2019,tay-khoo-chew-2017,ng-etal-2020}.) It is doubtful whether coverage, which is affected by so many parameters, is by itself a useful quantity to describe dimples performance.

\begin{figure}
\centering
\includegraphics[width=0.6\textwidth]{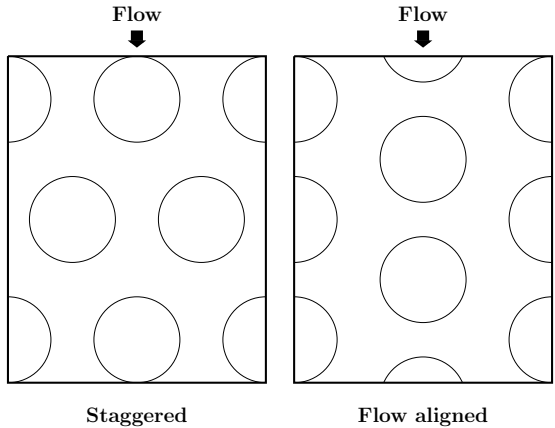}
\caption{Most popular dimples layout: staggered (left) and flow aligned (right).}
\label{fig:stag-alig}
\end{figure}

Moreover, dimples can be arranged either irregularly or regularly following a certain pattern. The two most widespread patterns are the staggered and the flow-aligned arrangements. Their definition is not unique. Often, the layout is referred to as staggered when the streamwise projection of one dimple overlaps with the following one, while it is called flow-aligned otherwise (see figure \ref{fig:stag-alig}). However, this definition, that corresponds to the most used arrangement, is not universally accepted. Prass et al. \cite{prass-etal-2019}, indeed, define the staggered arrangement as having the distance in spanwise direction from the centres of two adjacent dimples equal to half the distance between the centres of two non-adjacent dimples. Several additional patterns have been tested, e.g. the hexagonal one. 

\section{Do dimples work?}
\label{sec:overview}

Although in the last two decades a number of dimples-related contributions have appeared, many works claim that drag reduction is possible for certain geometries and flow conditions \cite{veldhuis-vervoort-2009, tay-2011, tay-khoo-chew-2015, vannesselrooij-etal-2016, tay-etal-2016, tay-lim-2017, tay-lim-2018, spalart-etal-2019}, whereas others only report drag increase \cite{tay-khoo-chew-2017, vancampenhout-etal-2018, prass-etal-2019, ng-etal-2020}. Notably, one work \cite{spalart-etal-2019} set out to specifically reproduce the experimental drag reduction results described in ref.\cite{vannesselrooij-etal-2016} with a state-of-the-art combined numerical/experimental study, and failed.

Such uncertain situation can be traced back to the lack of a generally accepted standard to measure drag and to compare different geometries among themselves and with the reference flat wall, since there are unavoidable differences when measuring drag in experiments and simulations, and in external (e.g. boundary layer) and internal (plane channel) flows. 
An additional reason explaining the scatter of available results consists in the still limited understanding of how dimples modify the surrounding flow field. 
Knowing the physics involved in drag reduction by dimples would be extremely useful in the optimization of all the several parameters involved. A description of the effect of the many geometrical parameters involved, and on the conjectures on the working mechanism of dimples are reported later in \S\ref{sec:parameters} and \S\ref{sec:physics} respectively.

We start by presenting an overview of the main results available in the literature, by focusing on the raw drag reduction information. 
They are reported in Table \ref{tab:comparison} that contains entries for the best drag reduction figure that could be extracted from each paper; when multiple dimple shapes are present, they are all considered. Drag change is simply defined here as $\Delta Drag = Drag_{dimples} - Drag_{smooth}$, where $Drag_{smooth}$ and $Drag_{dimples}$ are the (measured or computed) friction drag of the reference flat plate and the total drag of the dimpled plate, respectively. Negative values of $\Delta Drag$ thus correspond to drag reduction. Across the several studies, various definitions of the Reynolds number are used, particularly for internal flows. These have been all converted, whenever possible, to value of the bulk Reynolds number $Re_b$, by using the empirical Dean's correlation \cite{dean-1978}. Several other entries are also available in the Table, and will be defined and discussed throughout the text. 

\subsection{Experimental studies}
\label{sec:experiments}

The majority of the experimental studies carry out their tests in a wind tunnel and compare the drag measured on a flat plate with the drag measured on a dimpled plate. The flat/dimpled plate lies either on the upper or bottom wall, whereas the other wall of the wind tunnel is smooth. The plate is installed at a certain distance from the entrance section for the flow to become fully developed by the time it reaches the test section. A major difference among the various studies consists in the internal/external character of the flow.

The largest drag reduction, as observed in Table \ref{tab:comparison}, is a whopping 14\% found in the boundary layer experiment by Veldhuis \& Vervoort \cite{veldhuis-vervoort-2009} at the Technical University of Delft. The free-stream velocity was $7.5\ m/s$ and dimples were of circular shape. They found the staggered configuration to be more efficient in reducing drag than the flow-aligned one. Other boundary layer experiments carried out later by the same group at TU Delft reported a significantly smaller but still extremely interesting maximum drag reduction of 4\% \cite{vannesselrooij-etal-2016}, obtained at a Reynolds number based on the free-stream velocity $U_\infty$ and the dimple diameter $D$ of $Re_D \approx 40000$, which corresponds to a Reynolds based on the boundary layer thickness $\delta$ of $Re_\delta = 1500$. In this case, dimples are relatively large (in physical dimensions) shallow circular dimples, with a 50\% smaller coverage ratio than ref.\cite{veldhuis-vervoort-2009}. In a later study  \cite{vancampenhout-etal-2018}, they also measured a drag increase of 1\% for shallow dimples with different layouts at $Re_D \approx 63100$. Van Nesselrooij et al. \cite{vannesselrooij-etal-2016} presented what is described in ref.\cite{spalart-etal-2019} as a "very convincing experimental paper", studying different dimples configuration and finding that the best one consistently involves sparse (low coverage) and staggered dimples for the entire range of considered Reynolds numbers. They also focused on the importance of the depth of the dimples. When made dimensionless with the dimple diameter, shallower dimples are found to work better for each flow condition; however, when depth is compared to the boundary layer thickness, shallow dimples work better at low $Re$ but deep dimples are better at higher $Re$.

Another group that provided significant contribution to the dimples research thread is from the National University of Singapore, with Tay and colleagues. They performed experimental studies on a channel flow and reported up to 7.5\% drag reduction in ref.\cite{tay-lim-khoo-2019} for diamond-shaped dimples at a Reynolds number based on the bulk velocity $U_b$ and the channel semi-height $h$ of $Re_b \approx 30000$ and a layout with full coverage. Large drag reductions are measured also with other non-conventional dimple shapes \cite{tay-lim-2018}, such as the upstream-pointing teardrop at 6\%, or the downstream-pointing teardrop at 5\%, in a flow with $Re_b \approx 30140$ and $Re_b \approx 22270$, respectively. Conversely, the triangular shape was proved to always lead to drag increase \cite{tay-etal-2016}. Circular dimples were found to be less effective than diamond and teardrop shapes. A reduction of drag up to 2\% \cite{tay-2011} and 2.8\% \cite{tay-khoo-chew-2015} are found at $Re_b \approx 17500$ and $Re_b \approx 32100$ for different physical geometrical parameters of the dimple but with an identical layout and coverage ratio of 90\%. At $Re_b \approx 42850$ a drag reduction of 3.5\% is measured in ref.\cite{tay-lim-2018}. In ref.\cite{tay-khoo-chew-2015} they compare the same physical dimples and flow geometry by varying the coverage ratio and find that a dense layout with 90\% coverage performs better than a sparse one with 40\% coverage. They also compare two different dimple depths, measuring slightly higher drag reduction for deeper dimples. Finally, Tay \& Lim in ref.\cite{tay-lim-2017} experiment with shifting the point of maximum depth within the dimple along the streamwise direction, and measure the best performance of 3.7\% when the shift is $s=0.1D$ in the downstream direction.

\subsection{Numerical simulations}
\label{sec:simulations}

For drag reduction studies, numerical simulations need to resort to high-fidelity approaches, like Direct Numerical Simulation (DNS) and highly resolved LES (Large Eddy Simulation). Obviously, such simulations are not very practical for large-scale parametric studies, especially when the Reynolds number becomes large, since their unit computational cost rapidly increases with $Re$. The need for high-fidelity methods, the computational cost and the requirement to handle non-planar geometries are among the reasons why numerical studies for dimples are fewer than experiments. However, simulations (and DNS in particular, which avoids the need of turbulence modeling) are perfectly suited for such fundamental studies and provide us with the full information required to understand the working mechanism of dimples, by e.g. breaking up the drag changes into friction drag and pressure drag changes and by yielding a detailed and complete statistical characterization of the turbulent flow.

Circular dimples in a turbulent channel flow were studied with DNS for the first time in 2008 by Lienhart et al. \cite{lienhart-breuer-koksoy-2008}, who reported a drag increase of 1.99\% at $Re_b \approx 11000$. The same work contains also an experimental study of the same configuration, for which no drag changes were observed.

Ng et al. \cite{ng-etal-2020} at NUS performed one of the most interesting DNS studies, considering a turbulent channel flow at $Re_b=2800$ and examining different dimple geometries. They found that the classic circular dimple increases drag by 6.4\%, an amount that decreases to 4.6\% when the point of maximum depth is shifted downstream by $s=0.1D$. They also studied non-circular dimple shapes, obtaining this time a large drag reduction of 7.4\% for the diamond dimple, 4.9\% for the elliptical dimple and 3.1\% for the upstream-pointing teardrop dimple; the downstream-pointing teardrop dimple, instead, gave 0.1\% drag increase. 

Another recent numerical channel flow study is that by Tay and coworkers in \cite{tay-khoo-chew-2017}: they run a Detached Eddy Simulation (in which a baseline LES is complemented with a RANS model for the near-wall region) to replicate their own experimental study described in ref.\cite{tay-khoo-chew-2015}.  The DES yielded 1\% drag increase at $Re_b= 2830$, which does not confirm the experimental study and found drag increase for every case tested, whereas the experiments found smaller drag increase and even a slight drag reduction for a particular geometry. The suitability of DES for such drag reduction studies, however, remains dubious.

Prass et al. \cite{prass-etal-2019} published the only work in which an open channel is considered: with a LES they report a drag reduction of 3.6 \% at $Re_b \approx 6121$. They also considered two different configurations, finding that flow-aligned dimples perform better than staggered dimples.

There is only one DNS study for the boundary layer, i.e. the already mentioned work by Spalart \& at. \cite{spalart-etal-2019}, in which circular dimples at $Re_\delta = 30000$ were considered as the baseline geometry. They additionally studied the effect of the edge radius $r$, and found that with proper smoothing of this edge a drag reduction of -1.1\% is obtained, which descends from the combination of a -1.7\% reduction of friction drag, counterbalanced by a 0.6\% increase of pressure drag. 

\newcommand{\specialcell}[2][c]{%
  \begin{tabular}[#1]{@{}c@{}}#2\end{tabular}}
  
\begin{landscape}
\begin{table}
\caption{Summary of the main parameters and results of the literature. Columns report, in sequential order: 1. the reference and its acronym; 2. the numerical (DNS: direct numerical simulation, LES: large eddy simulation, DES: detached eddy simulation; FVM: finite volume method, SEM: spectral element method) or experimental approach; 3. the flow type: channel flow (CF), half-channel flow (HCF) or boundary layer (BL); 4. the value of the Reynolds number: $Re_b$ for CF and HCF, $Re_\delta$ for BL (other $Re$ definitions for CF are converted to $Re_b$ using the Dean's law); 5. the dimple shape: circular (Circ), triangular (Triang), diamond (Diam), elliptical (Ell), teardrop (Tear); upstream-pointing (Up), Downstream-pointing (Do); 6. spanwise width $D_z$ and streamwise length $D_x$, expressed as a fraction of the reference length $L$ (the channel half-height $h$ for CF and the boundary layer thickness $\delta$ for BL); for a circular dimple $D_z=D_x$, thus only one value is reported; 7. the dimple depth $d$; 8. the edge curvature radius $r$; 9. the curvature radius $R$ of the spherical cap; 10. the shift $s$ of the point of maximum depth; 11. the coverage ratio; 12. the type of layout: S:staggered, A: aligned, H:hexagonal; 13. the percentage drag change. "-" is used when some information required to compute the value is lacking. 
\label{tab:comparison}}
\newcolumntype{C}{>{\centering\arraybackslash}X}
\begin{tabularx}{1\linewidth}{CCCCCCCCCCCCC}
\toprule
\hline
\textbf{Article}   & \specialcell[t]{\textbf{Num/} \\ \textbf{Exp}}  &\textbf{Flow}   &\textbf{Re}    &\textbf{Shape} & \specialcell[t]{$\mathbf{D_z/L}$ \\ $\mathbf{(D_x/L)}$}  &$\mathbf{d/D_z \%}$   &$\mathbf{r/D_z}$   &$\mathbf{R/D_z}$    &$\mathbf{s/D_x}$   &\textbf{Cov$\%$} &\textbf{Layout} &$\mathbf{\Delta Drag \%}$  \\
\toprule
\multirow{2}*{\makecell{LBK-08 \\ \cite{lienhart-breuer-koksoy-2008}}} &Exp  &CF  &$\approx 10000$   &Circ  &$0.6$  &$5$ &- &- &$0$  &$22.5$ & A &$0$  \\   
                                                  &DNS (FVM)   &CF  &$10935$           &Circ  &$0.6$  &$5$ &- &- &$0$  &$22.5$ & A &$+1.99$  \\       
\midrule
\multirow{1}*{\makecell{VV-09 \\ \cite{veldhuis-vervoort-2009}}}      &Exp   &BL       &-                &Circ  &-      &$5$ &- &- &$0$  &$60$ &S &$-14$  \\
\\
\midrule                                                           
\multirow{1}*{T-11  \cite{tay-2011}}                    &Exp   &CF  &$\approx 17500$  &Circ  &$5$    &$5$ &$0.84$   &$1.68$ &$0$  &$90$  &S &$-2$  \\                                                                                                                                                                             
\midrule
\multirow{1}*{\makecell{TKC-15  \\ \cite{tay-khoo-chew-2015}}}          &Exp   &CF  &$\approx 32100$  &Circ  &$5$    &$5$  &$4.2$ &$8.45$ &$0$  &$90$   &S  &$-2.8$  \\  
\midrule
\multirow{1}*{\makecell{VVVS-16 \\ \cite{vannesselrooij-etal-2016}}}   &Exp   &BL       &$\approx 1500$   &Circ  &$26.67$ &$2.5$ &$0.5$ &$4.51$ &$0$  &$33.3$  &S &$-4$  \\ 
\midrule  
\multirow{3}*{\makecell{TLKJ-16 \\ \cite{tay-etal-2016}}}             &Exp   &CF  &$\approx 5625$  &Triang Up  &$- (4.67)$ &$5$   &- &- &$-0.5$  &-   &S  &$+4.8$  \\  
                                                  &Exp   &CF  &$\approx 5625$  &Triang Do  &$- (4.67)$ &$5$   &- &- &$+0.5$  &-   &S  &$+2.8$  \\  
                                                  &Exp   &CF  &$\approx 50350$ &Circ      &$5$    &$5$   &- &- &$+0.1$ &$90$   &S  &$-3.6$  \\                
\midrule              
\multirow{1}*{\makecell{TKC-17 \\ \cite{tay-khoo-chew-2017}}}         &DES (FVM)    &CF  &$\approx 2830 $  &Circ     &$5$    &$1.5$  &$0.84$ &$1.68$ &$0$  &$90$   &S  &$+1$  \\                              
\bottomrule
\end{tabularx}
\end{table}
\unskip
\end{landscape}

\begin{landscape}
\begin{table}
\newcolumntype{C}{>{\centering\arraybackslash}X}
\begin{tabularx}{1\linewidth}{CCCCCCCCCCCCC}
\toprule
\hline
\textbf{Article}   & \specialcell[t]{\textbf{Num/} \\ \textbf{Exp}}  &\textbf{Flow}   &\textbf{Re}    &\textbf{Shape} & \specialcell[t]{$\mathbf{D_z/L}$ \\ $\mathbf{(D_x/L)}$}  &$\mathbf{d/D_z \%}$   &$\mathbf{r/D_z}$   &$\mathbf{R/D_z}$    &$\mathbf{s/D_x}$   &\textbf{Cov$\%$} &\textbf{Layout} &$\mathbf{\Delta Drag \%}$  \\
\toprule              
\multirow{1}*{\makecell{TL-17 \\  \cite{tay-lim-2017}}}                &Exp   &CF  &$\approx 28600 $  &Circ    &$5$   &$1.5$ &- &- &$+0.1$  &$90$   &S  &$+1$  \\                                                                                                          
\midrule        
\multirow{3}*{\makecell{TL-18 \\ \cite{tay-lim-2018}}}                &Exp   &CF  &$\approx 42850$   &Circ    &$5$    &$5$   &- &- &$0$  &-   &S  &$-3.5$  \\                                 
                                                  &Exp   &CF  &$\approx 30140$   &Tear Up  &$5 (7.5)$    &$5$   &- &- &$+0.17$  &$84$   &S  &$-6$  \\   
                                                  &Exp   &CF  &$\approx 22270$   &Tear Dp  &$5 (7.5)$    &$5$   &- &- &$-0.17$  &$84$   &S  &$-5$  \\                                                                                           
\midrule   
\multirow{1}*{\makecell{VVVVS  \\-18 \cite{vancampenhout-etal-2018}}}     &Exp   &BL       &-                 &Circ    &-      &$2.5$     &$0.5$ &- &$0$  &-   &S/A/H  &$+1$  \\  
\\ 
\midrule                                                                               
\multirow{1}*{\makecell{SSSTPW \\ -19 \cite{spalart-etal-2019}}}           &DNS (FVM)    &BL       &$30000$           &Circ    &$1.33$ &-  &$2.03$  &-  &$0$  &- &S &$-1.1$  \\                                  
\midrule  
\multirow{1}*{\makecell{TLK-19 \\ \cite{tay-lim-khoo-2019}}}           &Exp   &CF  &$\approx 30000$   &Diam     &$5(10)$    &$5$         &- &- &$0$  &$99$   &S  &$-7.5$  \\                                                                                                 
\midrule         
\multirow{1}*{\makecell{PWFB \\ -19\cite{prass-etal-2019}}}            &LES (FVM)    &HCF    &$\approx 6121$  &Circ  &$5.7$ &$2.5$  &$1.5$ &$4.51$ &$0$  &- &S  &$-3.6$  \\      
\midrule                                               
\multirow{5}*{\makecell{NJLTK \\-20 \cite{ng-etal-2020}}}               &DNS (SEM)    &CF    &$2800$  &Circ       &$5$  &$5$   &$0.84$ &$1.68$ &$+0.1$  &$90.7$ &S  &$+4.6$  \\  
                                                  &DNS (SEM)    &CF    &$2800$  &Ell        &$5 (7.5)$  &$5$   &$0.84$ &$1.68$ &$0$  &$90.7$  &S  &$-4.9$  \\ 
                                                  &DNS (SEM)    &CF    &$2800$  &Tear Up        &$5 (8.75)$  &$5$   &$0.84$ &$1.68$ &$+0.21$  &$84.4$  &S  &$-3.1$  \\  
                                                  &DNS (SEM)    &CF    &$2800$  &Tear Dp        &$5 (8.75)$  &$5$   &$0.84$ &$1.68$ &$-0.21$  &$84.4$  &S  &$+0.1$ \\                                                
                                                  &DNS (SEM)    &CF    &$2800$  &Diam          &$5 (10)$  &$5$   &$0.84$ &$1.68$ &$0$  &$99.5$  &S  &$-7.4$  \\                                                                                     
\bottomrule
\end{tabularx}
\end{table}
\unskip
\end{landscape}

\section{How to design dimples?}
\label{sec:parameters}

Systematic studies which address the influence of each geometric parameter are lacking, so that the design of the optimal configuration to achieve the maximum drag reduction has not been identified yet. This Section describes the little we know, first in terms of the geometrical characteristics of the dimples and then in terms of their arrangement.

\subsection{The shape of the dimple}

\begin{figure}
\centering
\includegraphics[width=1.0\textwidth]{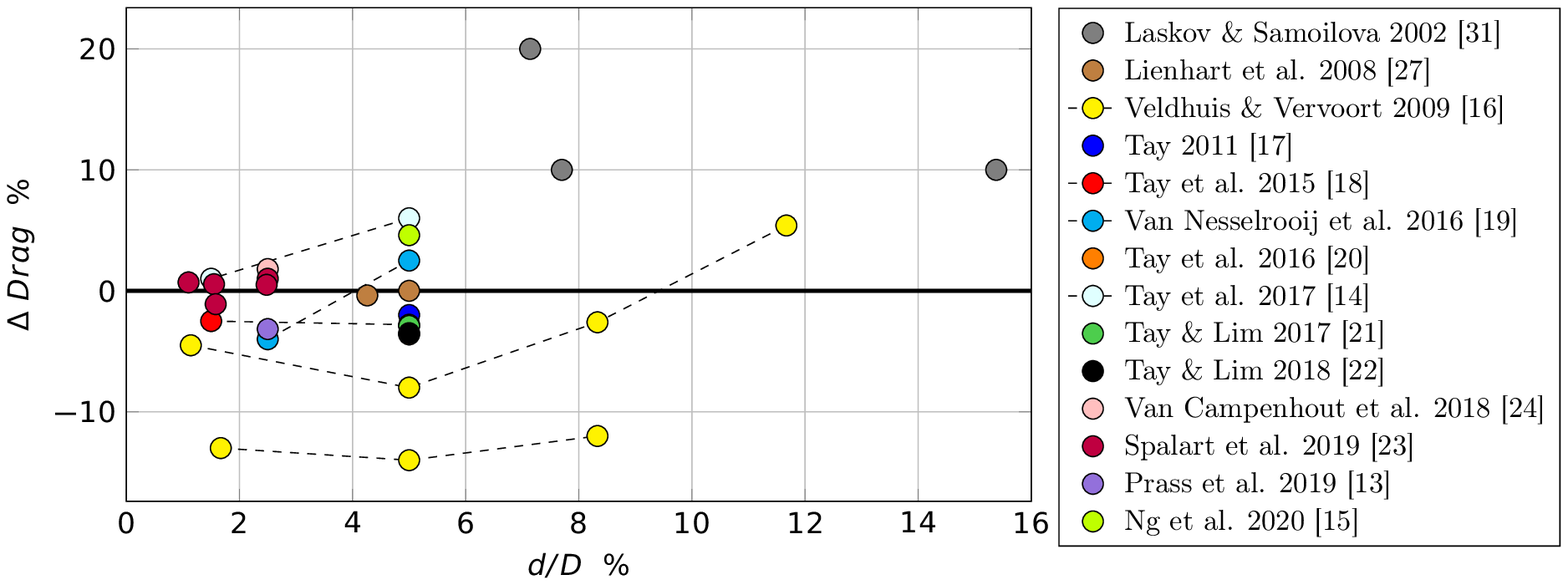} \\
\includegraphics[width=0.49\textwidth]{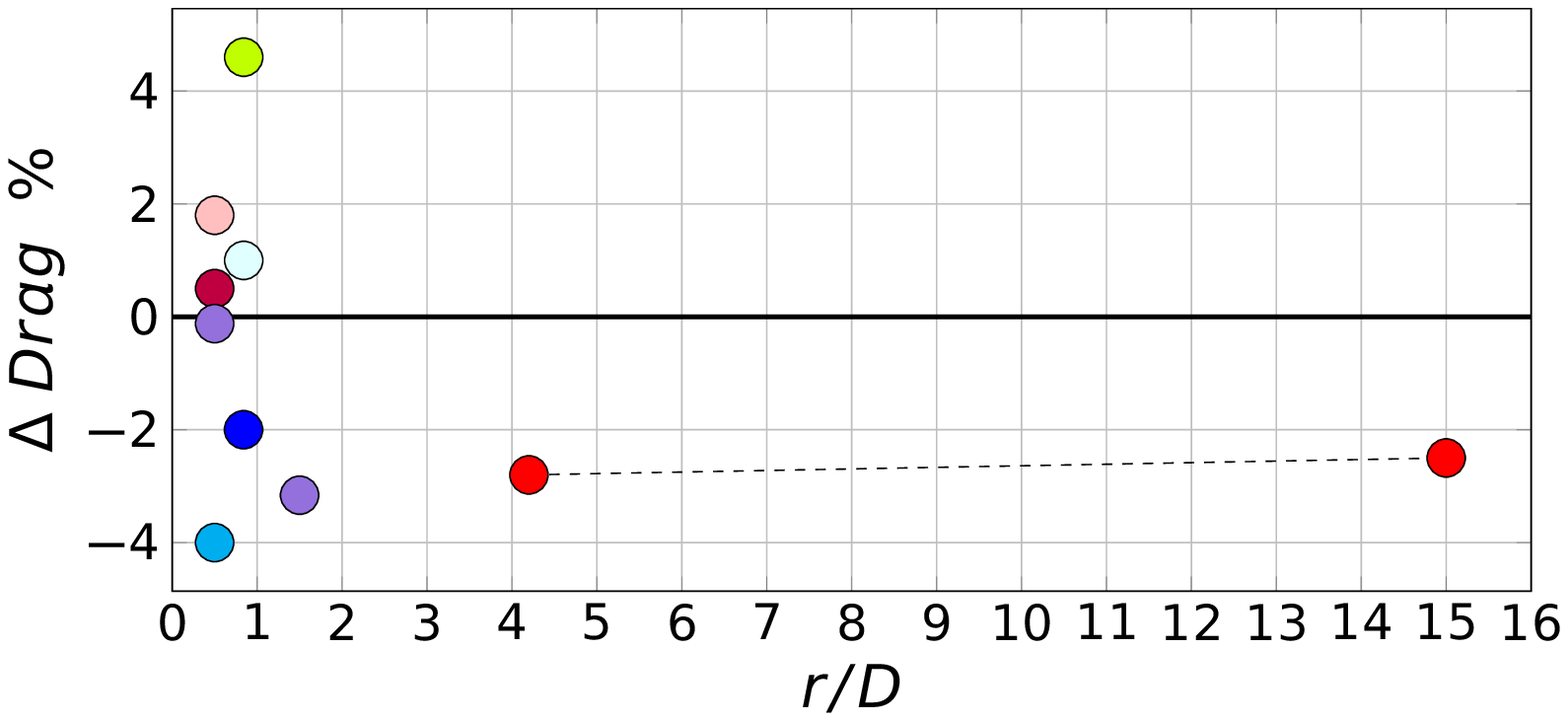}
\includegraphics[width=0.49\textwidth]{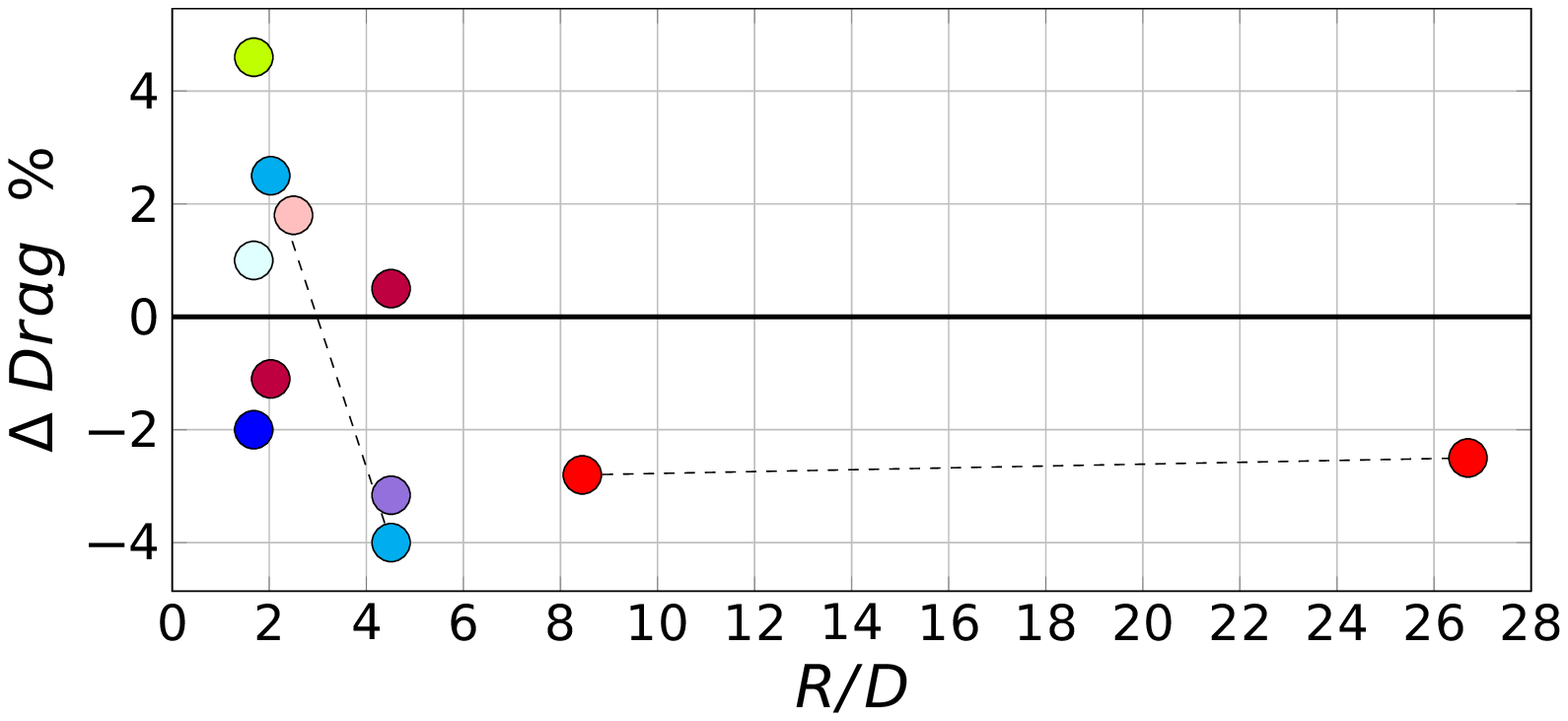}
\caption{Drag change obtained with circular dimples versus depth (top), edge curvature radius (bottom left) and radius of the spherical cap (bottom right). Dashed lines connect points for which only the parameter on the abscissa is changed.}
\label{fig:geom_par_dep}
\end{figure}
Figure \ref{fig:geom_par_dep} plots the drag change data measured by several works which adopted the baseline circular geometry. The percentage of drag change is shown as a function of the three independent geometrical parameters $d/D$, $r/D$ and $R/D$, after extracting from each publication the largest drag reduction (or the smallest drag increase). It should be noted that, in general, the various points correspond to simulations or experiments that differ for other, sometimes very important, design parameters. Dashed lines, instead, connect points for which only the parameter on the abscissa is changed.

The influence of $d/D$ on the drag change has been studied by several authors: previous research from heat exchange enhancement suggests the very reasonable idea that this is one of the key factors affecting drag. However, while the optimal $d/D$ is in the range $0.1-0.5$ for best heat exchange \cite{kovalenko-terekhov-khalatov-2010, tay-etal-2014}, several authors report that shallower dimples with $d/D < 0.1$ should be considered for reducing the overall drag, to avoid excess penalty from the ensuing pressure drag. 
Data are extremely scattered and clearly indicate that the drag change over a dimpled surface does not depend on the $d/D$ ratio alone. For example for $d/D=0.05$ Veldhuis and Vervoort \cite{veldhuis-vervoort-2009} report a drag reduction of almost 15\%, while Tay et al. \cite{tay-khoo-chew-2017} report a drag increase of approximately 6\%. The experimental measurements from ref.\cite{veldhuis-vervoort-2009} are for a turbulent boundary layer over a dimpled surface with coverage ratio of 60\% at a free-stream velocity in the range $0-29\ m/s$; the results from ref.\cite{tay-khoo-chew-2017} are from a Detached Eddy Simulation of a turbulent channel flow at $Re_b \approx 3000$ with a coverage ratio of 90\%. 

It is reassuring, though, to see --- at least in some of the datasets where data points are connected by dashed lines --- a local optimum for intermediate depths, since it is reasonable to expect zero drag changes for $d \to 0$ and an increase of drag as for standard $k$-type roughness for $d \to \infty$.
With the other parameters unchanged, Tay et al. \cite{tay-khoo-chew-2017} and Van Nesselrooij et al. \cite{vannesselrooij-etal-2016} agree on observing a decrease of performance with increasing $d/D$ (in the range $0.015 < d/D < 0.05$), although at a different rate; within the same $d/D$ range, Tay et al. \cite{tay-khoo-chew-2015} and Veldhuis and Vervoort \cite{veldhuis-vervoort-2009} measured a slight increase of drag reduction performance with increasing $d/D$. 
The latter study was extended up to $d/D =0.12$, finding that for $d/D>0.05$ the overall drag increases with $d/D$.

The curvature radius $r$ at the edge of the dimple is meant to mitigate the negative effects of pressure drag, by preventing or decreasing flow separation. The second panel of figure \ref{fig:geom_par_dep} shows that also in this case data are highly scattered: for $0.5 \lessapprox r/D \lessapprox 1.5$ the achieved drag change ranges between -4\% \cite{vannesselrooij-etal-2016} and 4.8\% \cite{ng-etal-2020}. The experiments of Tay et al. \cite{tay-khoo-chew-2015} at $Re_b \approx 32100$ show that after a certain value, i.e. $r/D > 4$, the influence of the edge curvature on the drag change is minimal. Spalart et al. \cite{spalart-etal-2019} performed a DNS of a turbulent boundary layer, with a Reynolds number (based on the boundary layer thickness) of $Re_\delta = 7.5 \times 10^3 $ and $ Re_\delta=3 \times 10^4$ and considered $r/D=0.5$. Their data points are not plotted in figure \ref{fig:geom_par_dep}, since their paper does not contain enough information to quantify $r$. However, they confirmed that smoothing the dimple rim is beneficial.

A scattered picture is also obtained when data are plotted against the $R/D$ ratio, as shown in the third panel of fig.\ref{fig:geom_par_dep}, confirming again that for this configuration a single geometrical parameter is unable to fully characterize the influence of the dimples on the flow.

The experiments of Tay \& Lim \cite{tay-lim-2017} and the numerical simulations of Ng et al. \cite{ng-jaiman-lim-2018, ng-etal-2020} agree on the indication that the downstream shift $s$ is beneficial, for a wide range of Reynolds numbers, with the best effect observed when $s=0.1D$ in the downstream direction. When instead the shift is in the upstream direction, i.e. $s<0$, drag increases rapidly. 
It should be mentioned that the Reynolds number of the simulations ($Re_b=2800$) is somewhat lower than the lowest Reynolds number of the experiments ($Re_b \approx 4300$). Tay \& Lim \cite{tay-lim-2017} claim that a $0.2D$ downstream shift is equivalent to the axisymmetric case at $Re_b=7000$ with a drag increase of 1.5\%, because the lower drag obtained by the reduced flow separation at the shallower upstream wall is compensated by the higher drag of the flow impinging on the steeper downstream wall. Ng et al. \cite{ng-jaiman-lim-2018, ng-etal-2020}, who can take advantage of DNS to break down the total drag into friction and pressure contributions, find that friction drag is almost unaffected by a downstream shift, since it does not affect the reattachment point.

When it comes to alternative shapes, triangular dimples were considered by Tay et al. \cite{tay-etal-2016}. In their experiment they machined dimples with the bottom surface sloping up from the deepest point at the triangular vertex towards the base of the triangular depression to meet the flat channel surface, hence producing the negative of a wedge. A larger drag was obtained for both upstream- and downstream-pointing triangles, for the whole range of tested Reynolds numbers, i.e. $ 5180 \le Re_b \le 28600$. Moreover, for the downstream-pointing triangle the drag increase is nearly constant with $Re$, whereas for the upstream-pointing triangle the drag increase grows with $Re$.
Tay \& Lim \cite{tay-lim-2018} studied the teardrop dimple and measured drag reduction for both the upstream- and downstream-pointing teardrops, for $4500 \le Re_b \le 44000$, with the former yielding up to 6\% drag reduction and the latter up to 5\%. Tay et al. \cite{tay-lim-khoo-2019} studied the diamond dimple and measured drag reduction up to 7.5\%. More recently, Ng et al. \cite{ng-etal-2020} compared in a numerical study the circular, elliptical, teardrop and diamond dimples in a turbulent channel flow, reporting drag reduction of 4.9\% for the elliptical dimple, 3.1\% for the upstream-pointing teardrop, and 7.4\% for the diamond dimple.

\subsection{The arrangement of the dimples}

When it comes to the spatial arrangement of dimples on the surface, once the other parameters are fixed, the staggered configuration leads to lower drag compared to the flow-aligned one \cite{veldhuis-vervoort-2009, vannesselrooij-etal-2016, vancampenhout-etal-2018, spalart-etal-2019}, a fact that explains why the staggered configuration is the most adopted one. Van Nesselrooij et al. \cite{vannesselrooij-etal-2016} found 3\% of drag increase for flow-aligned dimples and up to 4\% drag reduction for staggered dimples with the same geometrical parameters, coverage and Reynolds number. Spalart et al. \cite{spalart-etal-2019} found drag increase for both configurations, but the drag increase of the flow-aligned dimples was almost twice that of the staggered dimples.

Lashkov and Samoilova \cite{lashkov-samoilova-2002} and Van Campenhout et al. \cite{vancampenhout-etal-2018} considered the drag change also for other, non-standard arrangements. The former study found a large drag increase (up to 50\%) when an hexagonal dimple layout is used. The latter study showed a constant drag increase of about 1\% for each of the several considered layouts.

The effect of coverage ratio was considered by Tay et al. \cite{tay-2011, tay-khoo-chew-2015}, who compared in a channel flow circular dimples with 40\% and 90\% coverage, and found that higher coverage enhances the (positive or negative) effects of the dimples. 
Van Nesselrooij et al. \cite{vannesselrooij-etal-2016} experimentally studied the effect of coverage in a boundary layer. They found that a 90\% coverage yields drag increase for a wide range of $Re$, whereas 33.3\% coverage always yields drag reduction within the same Reynolds number range. Performance of both layouts are found to improve by increasing $Re$. Spalart at al. \cite{spalart-etal-2019} in their boundary layer DNS compared the two coverage ratios, and observed about 1\% of drag reduction for the lower coverage, and 2\% of drag increase for the higher one.

\section{How do dimples work?}
\label{sec:physics}

The uncertainties on the true effectiveness of dimples in reducing turbulent drag are accompanied, perhaps unsurprisingly, by a limited understanding of the physics involved. Thanks to the several experimental and numerical works carried out so far, some ideas and hypotheses exist, but consensus is lacking. We describe below two prevailing descriptions of how dimples interact with the overlying turbulent flow.

\subsection{Self-organized secondary tornado-like jets}

The first attempt at explaining drag reduction by dimples is due to Kiknadze et al. \cite{kiknadze-gachechiladze-barnaveli-2012}, who based their explanations uniquely of video records and photographs, even though similar observations were already put forward in previous numerical \cite{veldhuis-vervoort-2009} and experimental \cite{kovalenko-terekhov-khalatov-2010} studies. According to ref.\cite{kiknadze-gachechiladze-barnaveli-2012}, whose authors are affiliated with the Research and Production Centre “Tornado-Like Jet Technologies” in Moscow, the action of dimples can be explained by a so-called tornado-like jet self-organization. In plain words, this is how the flow organizes itself and develops over the double-curvature concavity of a dimple. The flow coming from an upstream flat portion accelerates at the leading edge of a circular dimple, and is lifted off from the surface while trying to follow the curved wall, leading to a reduction of skin-friction drag in the fore half of the dimple. After the streamwise midpoint, the flow converges towards the midline to eventually meet the flat wall past the trailing edge, and the skin friction increases again. Although the skin friction reduction in the fore half might outweigh the increase of the aft half, the recessed geometry of the dimple introduces an additional pressure drag component: hence, to achieve drag reduction the net reduction of the skin-friction drag needs to be larger than the increase of pressure drag. It should be observed, though, that this description is not directly addressing the insurgence of drag reduction, but only constitutes an attempt to draft a simplified description of the local flow modifications induced by the dimple.  

If dimples are deep enough, their steep walls make the flow prone to separation on the upstream part of the recess, with creation of spanwise vorticity and recirculation. 
The flow reversal has a positive effect on the drag, causing negative skin friction in the first portion of the dimple. 
When the flow reattaches, a strong impingement of the flow on the rear slope of the dimple produces a locally high skin friction. 
Moreover, flow separation obviously causes a large increase of pressure drag which could cancel out the positive effect of the skin friction drag. 
To avoid separation and the consequent increase of pressure drag, more efficient shapes than the classical circular one are used. 
Shifting downstream the point of maximum depth of the dimple alters the wall slopes, and affects the total drag by changing pressure drag, whereas the friction drag tends to remain unchanged \cite{ng-etal-2020}. 
A (moderate) downstream shift minimizes the negative effects of separation, and offers lower drag than the standard circular geometry.
However, the shift does not significantly affect the location of the reattachment point, except for very large shifts, for which flow reversal may be entirely suppressed, but at the cost of an intense impingement onto the steeper rear wall which negatively affects the drag. 

Non-circular dimples induce different drag changes \cite{ng-etal-2020}. Flow separation and flow reversal are not observed for elliptical, upstream-pointing and diamond dimples, leading to a lesser drag compared to the smooth wall. This can be attributed to the gentler upstream slope and to the longer, more streamwise-aligned leading edge.
Other studies which do not report flow reversal even for the circular shape are \cite{vannesselrooij-etal-2016, spalart-etal-2019}; they measure a maximum drag reduction of 4\% and 1.1\%, respectively. Tay et al. \cite{tay-khoo-chew-2017} observe flow separation for circular dimples in the whole range of tested flow conditions for $d/D = 0.05$, but not for $d/D=0.015$; however, they measure drag increase in all the tested cases.

\subsection{Spanwise forcing}
A more recent conjecture on the mechanisms by which dimples attain drag reduction has been put forward independently by the two groups at TU Delft \cite{vannesselrooij-etal-2016} and NUS \cite{tay-khoo-chew-2015}.
Flow visualisations indicate that, near the wall, streamlines coming in straight from a flat surface bend towards the dimple centerline in the upstream portion of the recess, then bend away from it in the downstream portion, thus creating a converging-diverging pattern (see for example \cite{tay-etal-2014}). 
Such meandering implies a spanwise velocity distribution with changing sign across the dimple length \cite{vannesselrooij-etal-2016, vancampenhout-etal-2018}, and a consequent alternating streamwise vorticity \cite{tay-khoo-chew-2017} since the spanwise velocity remains confined very near to the wall. Ref. \cite{vannesselrooij-etal-2016} reports an average spanwise velocity of about 2--3\% of the free-stream velocity for a boundary layer; ref.\cite{tay-khoo-chew-2017} measured a maximum spanwise velocity in the range 3.5--8\% of the centerline velocity in the channel. Spalart et al. \cite{spalart-etal-2019} also detected in their DNS study a spanwise motion, although weaker in intensity. 

\begin{figure}
\centering
\includegraphics[width=0.7\textwidth]{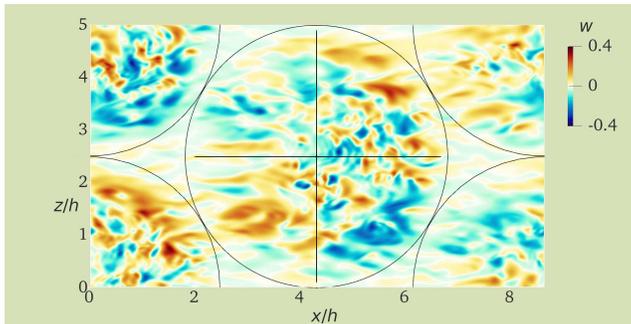}
\caption{Instantaneous spanwise velocity component $w$ on a wall-parallel plane at $y^+= 1.3$ from the flat part of the wall. Lengths and velocities are made dimensionless with $h$ and $\mathsf{U_b}$. The velocity field is computed by DNS for a circular dimple, which actually yields drag increase.}
\label{fig:spanwise}
\end{figure}

Figure \ref{fig:spanwise} depicts an instantaneous spanwise velocity field over a circular dimple, taken from one of our DNS simulations of a turbulent channel flow over circular dimples (see Appendix \ref{sec:appendix} for computational and discretization details of our simulations). An alternating spanwise velocity pattern is clearly visible, supporting the idea that the dimple creates a velocity component in the spanwise direction and bends the streamlines in a converging-diverging behaviour. The instantaneous values are very large, up to 40\% of the bulk velocity.

The alternate spanwise velocity resembles the spanwise-oscillating wall \cite{jung-mangiavacchi-akhavan-1992}, an active technique for the reduction of turbulent friction drag, where the wall oscillates in time in the spanwise direction. In the oscillating-wall control, the spanwise  velocity component at the wall $w_w$ is prescribed as a function of time as:
\begin{equation}
w_w(t) = A \sin \left( \frac{2 \pi}{T} t  \right),
\end{equation}
where $A$ is the amplitude of the oscillation and $T$ is its period.
The oscillating wall produces very large reductions of friction drag, although at a significant energy cost. Its detailed performance is determined by the control parameters $A$ and $T$; Quadrio \& Ricco \cite{quadrio-ricco-2004}, after a careful DNS study, identified the link between the value of parameters and the obtained drag reduction. They found an optimum value for the oscillating period of $T^+ \approx 100$, whereas drag reduction monotonically improves with the amplitude (albeit the energy cost of the control rises faster as $A^2$).
Dimples could be considered as a passive implementation of the spanwise-oscillating wall. Van Campenhout et al. \cite{vancampenhout-etal-2018} measured the analogous parameters and defined a period $T$ and a maximum spanwise velocity $w_{max}$ of a fluid particle, averaging over a selected region of the domain. In the oscillating wall, it is known \cite{quadrio-sibilla-2000} that the time-averaged mean spanwise velocity profile coincides with the laminar solution of the Stokes second problem. Ref.\cite{vancampenhout-etal-2018} assumes the same to hold for the flow over dimples, thus deriving an analogous value for the amplitude. For their dimples with $d/D=0.025$, they found $T^+=135$ and $A^+=0.74$. Data from ref.\cite{quadrio-ricco-2004} do not contain information for such small amplitudes, but an extrapolation leads to a drag reduction of about 4\% for this combination of parameters: a value that closely resembles the measurement of 3.8\% from ref.\cite{vannesselrooij-etal-2016}. 

It should be noted, first, that a closer analogy should be made between this interpretation of the dimples working mechanism and the spatially modulated spanwise forcing introduced by Viotti et al. \cite{viotti-quadrio-luchini-2009}. However, in that paper it is shown how temporal and spatial oscillations can be easily converted one into the other by using a suitable convective velocity scale at the wall. There are, of course, obvious differences between data collected by Quadrio \& Ricco for a turbulent channel flow at $Re_\tau=200$ or $Re_b=3173$ and the dimple experiments described in refs.\cite{vannesselrooij-etal-2016, vancampenhout-etal-2018} for a boundary layer at $Re_\delta=1226$ (the limited information provided in these references precludes computing the value of the friction Reynolds number). 

Other important concepts to be aware of when trying to draw such a parallel is that, with the oscillating wall, a minimum spanwise velocity is required for the active technique to produce its effects: this threshold value $A_{th}^+$, that needs to be of the order of the natural fluctuations of spanwise velocity in the near-wall region, is quantified in ref.\cite{quadrio-ricco-2004} as $A_{th}^+=1$, i.e. similar or larger than the dimples-induced spanwise velocity as determined in \cite{vancampenhout-etal-2018}. Finally, and definitely most important, with a flat wall, even in presence of spanwise forcing, one should be only concerned with friction drag, whereas with dimples both viscous and pressure drag come into play.

\section{How to set up a proper comparison?}
\label{sec:analysis}
Measuring (small) changes in aerodynamic drag is not trivial, especially in the turbulent regime, regardless of the numerical or experimental nature of the analysis. Studies employ a variety of approaches, where simulations and experiments presents different approaches and different challenges.

Nowadays, whenever we need to compare the drag of a reference flat surface with that of a rough surface, we are aware of the subtlety of the measurement, of the importance to carefully define and control the Reynolds number of the experiment, to discriminate between internal and external flows, and in general to correctly define the equivalent "flat wall" flow to compare with. In this final Section, we will discuss some of these topics, trying to call the reader's attention to the logical steps that should be followed when designing a meaningful experimental or numerical campaign.

\subsection{Measurement of the drag (difference)}
All the available studies measure the drag difference $\Delta Drag$ by separately measuring the drag forces $Drag_{smooth}$ and $Drag_{dimples}$. As recently discussed in ref.\cite{vannesselrooij-etal-2022} in the context of the description of their novel experimental setup devoted to such measurements, various approaches are available. The simplest among them measure the local friction, and as such are unable to yield satisfactory results for the drag, because the friction contribution to the drag force over a dimpled surface depends on the position, and the same holds for the pressure component. Hence, in an experiment one has to either resort to measuring the drag force with a balance, a challenge by itself owing to the small forces involved, or to deduce the force from the pressure drop across two sections, as done for example by Gatti et al. in \cite{gatti-etal-2015}. With dimples, both approaches have been used. Information about the shear stress was extracted from boundary layer momentum loss in ref. \cite{lienhart-breuer-koksoy-2008}. Direct measurement of the drag through a force sensor was employed in refs.\cite{veldhuis-vervoort-2009, vannesselrooij-etal-2016, vancampenhout-etal-2018}. This type of measurement may be affected by uncertainty and accuracy problems: forces are small, and blurred by spurious contributions, and the experimental setup must be designed and run with extreme care.

In the case of numerical experiments, only the DNS approach provides the required accuracy that is not embedded e.g. in RANS models, constructed and tuned for canonical flows and hence incapable to deal with drag reduction in a quantitatively accurate way. Once DNS is used, two equivalent options are available to compute the drag in internal flows. One possibility is the calculation of the (time-averaged value of) the friction drag and the pressure drag separately, employing their definition as surface integrals of the relevant force component.  
In alternative, the (time-averaged value of) the pressure drop between inlet and outlet informs of the total dissipated power, and thus leads to the total drag. This is feasible both in simulations and experiments. Tay and colleagues \cite{tay-2011, tay-khoo-chew-2015, tay-etal-2016, tay-khoo-chew-2017, tay-lim-2017, tay-lim-2018} in fact compared the mean streamwise pressure gradients of both the two flat sections upstream and downstream of the dimpled test section with the mean streamwise pressure gradient within the test section, employing static pressure taps.

Experience accumulated in riblets research, however, tells us that the riblets community obtained its first fully reliable dataset when D.Bechert in Berlin developed on purpose a test rig, the Berlin oil channel \cite{bechert-etal-1992}, where the measured quantity was directly the drag difference: targeting the quantity of interest, i.e. the drag difference under identical flow conditions,  instead of relying on the difference between two separately measured drag forces was key to improve accuracy and reliability.

\subsection{The Reynolds number}

Dynamic similarity is a well known concept in fluid mechanics, and enables meaningful comparative tests provided the value of the Reynolds number is the same. The true question is to understand {\em which} Reynolds number should be kept the same. The Reynolds number is defined as the product of a velocity scale $U$ and a length scale $L$, divided by the kinematic viscosity $\nu$ of the fluid. While e.g. in an experiment the precise measurement of $\nu$ might be difficult, its meaning is unequivocal. Choosing $U$ and $L$, instead, presents more than one option. 

For the velocity scale $U$, dimples do not lead to specific issues. While for a zero-pressure-gradient boundary layer over a flat plate the use of the external velocity $U_\infty$ sounds reasonable, for internal flows like the plane channel flow one has to choose among the bulk velocity $U_b$, the centerline velocity $U_c$ and the friction velocity $u_\tau$. The choice of reference velocity has been already discussed in the context of skin-friction drag reduction \cite{hasegawa-quadrio-frohnapfel-2014}: provided drag reduction is not too large, and the flow is far enough from laminarity, choosing $U$ is not critical and should not be regarded as a major obstacle.

For the length scale $L$, instead, the situation is different, as dimples themselves contain one or more length scales that could be used in the definition of $Re$. For example, to avoid the ambiguity implied by the definition of the origin for the wall-normal coordinate, Van Nesselrooij et al. \cite{vannesselrooij-etal-2016} and Van Campenhout et al. \cite{vancampenhout-etal-2018} for their boundary layer experiments decided to define a Reynolds number based on the diameter of their circular dimple. Naturally, achieving the same $Re$ based on flow velocity and dimple diameter is not enough to guarantee dynamic similarity in two different flows.

\begin{figure}
\includegraphics[width=0.9\textwidth]{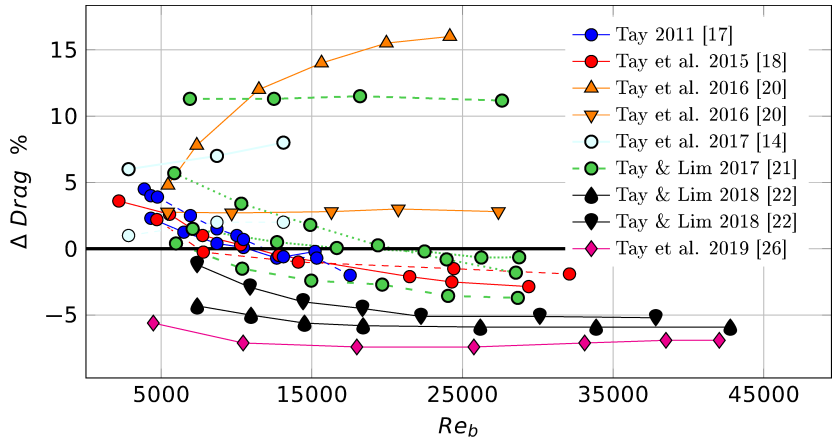}
\caption{Drag change versus bulk Reynolds number $Re_b$.}
\label{fig:Re}
\end{figure}

By isolating all the data sets for which a value for the bulk Reynolds number $Re_b$ is given (either explicitly or deduced from equivalent information), and putting together the reported drag changes, one obtains the picture reported in figure \ref{fig:Re}. Besides showing both drag reduction and drag increase, drag changes exhibit every possible trend with $Re_b$: increasing, decreasing, constant or nearly constant, and non-monotonic with either a maximum or a minimum at intermediate $Re_b$. Without excluding additional possible causes, this can be attributed to the host of parameters that are not kept identical across the dataset, besides the Reynolds number, and stresses once more the importance of experiments where only one parameter is changed at a time. 

\begin{figure}
\centering
\includegraphics[width=1\textwidth]{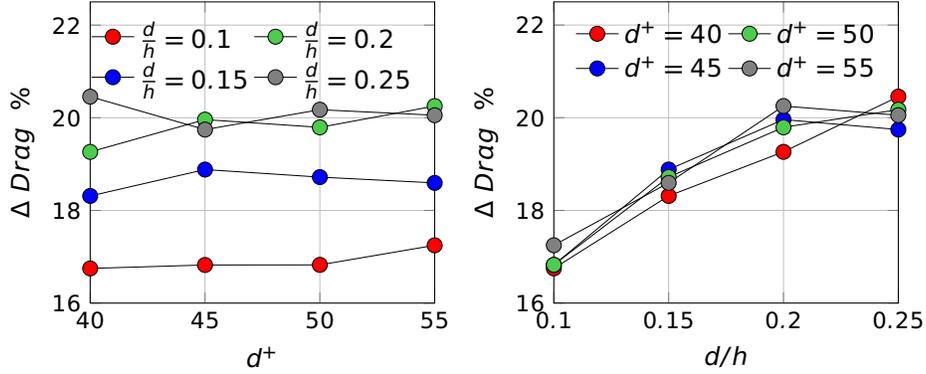}
\caption{Present simulations, circular dimples at various sizes and Reynolds numbers with $2690 \le Re_b \le 10450$. Left: drag changes vs dimple depth in inner units. Right: drag changes vs dimple depth in outer units.}
\label{fig:deep-dimples}
\end{figure}

In a turbulent wall flow, the Reynolds number is an essential ingredient to define the proper scaling of important quantities, say the total drag change. If for example only the dimple depth $d$ is varied, its value can be set in wall units ($d^+$) or in outer units ($d/h$), and, if the Reynolds number is also changed, various combinations for $d^+$ and $d/h$ become possible. It is the flow physics which dictates what scaling works best at collapsing results. 
We have performed two sets of DNS simulations (see \S\ref{sec:appendix} for details) to understand the scaling of drag changes induced by circular dimples when only their dimensions are changed but its shape is preserved. We have fixed the values of $d/D$ and $r/R$, the value of the depth $d$ (either in inner $d/h$ or outer $d^+$ units) has been varied, and all the other parameters did vary accordingly, as prescribed by equation \eqref{eq:geom_rel}. 

Figure \ref{fig:deep-dimples} plots the results and shows that drag changes (in this specific case, drag increases) appear to follow an outer scaling: all the data points collapse onto a single curve when drag changes are plotted against $d/h$. This is an expected result, as these dimples are rather deep, and thus somehow akin to a large-scale $d$-roughness \cite{jimenez-2004}, where the large cavities basically destroy the near-wall layer, i.e the only region where inner scaling would make sense.

\subsection{The equivalent flat wall}

The comparison between flat and dimpled wall can be set up for internal or external flows. The latter, which may be less convenient in numerical simulations owing to their non-parallel nature, present a sensible advantage in this context, since drag and its related changes have simply to be computed for the same plate immersed in the same external velocity, and a reduced drag force is unequivocally advantageous. For internal flows, however, the non-planar dimpled wall brings up the problem of properly defining the location of the equivalent flat wall and, in general, of setting up the comparison properly. 

\begin{figure}
\includegraphics[width=\textwidth]{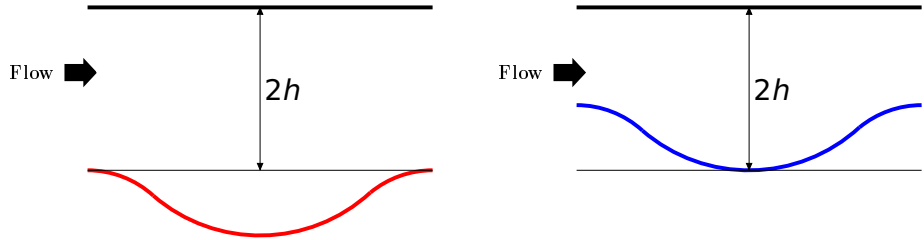}
\caption{A dimpled wall and two different, equivalent flat channels. The red/blue lines indicate the dimple profile. Left: the channel height $2h$ goes from the top wall to the dimple tip; right: the channel height $2h$ goes from the top wall to the dimple lowest point.}
\label{fig:equiv-channel}
\end{figure}

As shown schematically in fig.\ref{fig:equiv-channel}, for a channel flow, for example, a certain definition of the reference flat wall impacts the reference length $h$ and, eventually, changes the value of the Reynolds number of the flow to compare with. The reference wall might be placed on the flat surface among dimples, on the position of lowest elevation in the cavity, on the average height of the dimpled surface, etc., leading to different flow volumes. 


To properly account for this effect, let us start from the usual definition of the bulk Reynolds number $Re_b = U_b h / \nu$, where $h$ is a reference length (e.g. half the witdh of the flat channel) and $\nu$ is the kinematic viscosity. Once the cross-sectional area $A(x)$ of the dimpled channel changes along the streamwise direction, the bulk velocity $U_b$, defined as an average velocity across the section, becomes itself a streamwise-dependent function:
\begin{equation}
  U_b(x)= \frac{1}{A(x)}\int_{A(x)} u(\mathbf{x}) \text{d}A .
\end{equation}
We thus replace this definition with a volume average, and define a new bulk velocity $\mathsf{U_b}$ as an average over the volume to obtain a streamwise-independent quantity:
\begin{equation}
  \mathsf{U_b} = \frac{1}{V} \int_V u(\mathbf{x}) \text{d}V .
\end{equation}
Note that the two quantities $U_b$ and $\mathsf{U_b}$ coincide for a flat wall. A comparison at same flow rate requires that the volumetric flow rate
\begin{equation}
  Q = \int_{A(x)} u(\mathbf{x}) \text{d}A = \frac{1}{L_x} \int_0^{L_x}\int_{A(x)} u(\mathbf{x})\text{d}A\text{d}x = \frac{1}{L_x} \int_V u(\mathbf{x}) d V = \frac{V}{L_x} \mathsf{U_b}
\end{equation}
is the same for the flat and dimpled channels, provided the streamwise length $L_x$ of the channel is the same. This implies that $V_f \mathsf{U_{b,f}} = V_d \mathsf{U_{b,d}}$, where the subscripts $\cdot_f$ and $\cdot_d$ refer to quantities measured in the flat and dimpled channel respectively. In the end, the bulk velocity in the dimpled channel (and the bulk Reynolds number) need to be changed by multiplication of a factor given by the volume ratio:
\begin{equation}
  \mathsf{U_{b,d}} =\frac{V_f}{V_d} \mathsf{U_{b,f}};
  \qquad
  Re_{b,d} = \frac{V_f}{V_d} Re_{b,f}.
\end{equation}

The numerical value of $Re_b$ is thus affected by the choice of the equivalent flat channel. For example, the equivalent flat channel might go from the top wall to the lowest point of the dimple, and $Re_{b,d}>Re_{b,f}$. In contrast, if the equivalent channel goes from the top wall to the tip of the dimple, $Re_{b,d}<Re_{b,f}$. The two bulk Reynolds numbers end up being the same only when the volume is preserved in the reference and dimpled channels (i.e. the equivalent flat channel is located at the average dimple height).

If the comparison is carried out by DNS, one conveniently measures the time-averaged value $\overline{f}$ of the spatially uniform volume force $f$ required to maintain a constant flow rate at each time step. This volume force is interpreted as $f = \Delta P / L_x$, where $\Delta P$ is the pressure drop along the channel. The proper measure of the drag change is:
\begin{equation}
  \Delta Drag = \frac{ V_d \overline{f_d} - V_f \overline{f_f} }{ V_f \overline{f_f} } = \frac{ V_d/V_f \overline{f_d} - \overline{f_f}}{\overline{f_f}};
\end{equation}

Therefore, the change of the fluid volume has to been considered also when measuring the drag change in the controlled case.

\begin{figure}
\centering
\includegraphics[width=0.6\textwidth]{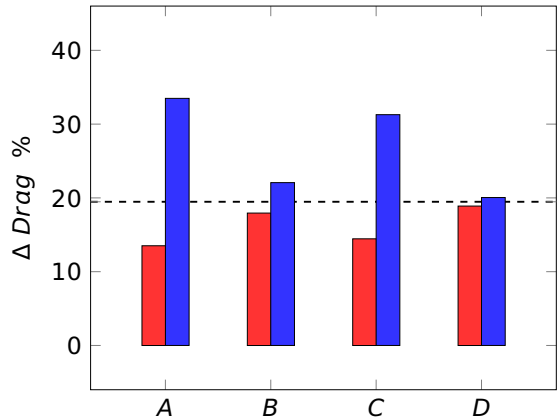}
\caption{Drag changes, measured by DNS, for circular dimples with $d/h=0.25$ at $Re_b \approx 2800$. Red/blue bars express drag changes when the equivalent channel defines $2h$ as the distance between the top wall and the top/bottom of the dimple (color code is the same of figure \ref{fig:equiv-channel}). Case A: comparison at the same $Re_b$, $\Delta Drag$ computed without accounting for the volume ratio. Case B: as case A, but $\Delta Drag$ is corrected with the volume ratio. Cases C and D are like cases A and B, but the comparison is made at the same flow rate.}
\label{fig:origin}
\end{figure}

Figure \ref{fig:origin} exemplifies the consequences of neglecting these considerations. These are certainly exaggerated by the choice of working with a dimple configuration that causes a large change of drag. However, the relative differences are major; neglecting such considerations would most certainly hinder the true ability of dimples to alter skin-friction drag.

\subsection{The drag reduction metrics}

In closing, we mention a final methodological issue, that affects drag reduction measurements for dimples, riblets, and roughness at large: the proper metrics to express it. It is customary to express drag reduction as (percentage) changes in the skin-friction coefficient at a given $Re$; unfortunately, the coefficient itself contains a dependence on the Reynolds number already for the flat wall case, thus making it impossible to rely on percentage changes for a robust assessment of the drag change properties of a given surface. The complete information would be the $(\Delta Drag, Re)$ pair. In alternative, the proper metric for expressing drag reduction is the vertical shift of the logarithmic portion of the mean streamwise velocity profile expressed in viscous units.

This is a known concept for roughness \cite{jimenez-2004} as well as riblets \cite{luchini-1996, spalart-mclean-2011}, and also extends to some active flow control strategies \cite{gatti-quadrio-2016}. As long as the direct effect of the roughness remains confined within the buffer layer of the turbulent flow, it can be translated into an upward shift $\Delta U^+$ of the logarithmic velocity proﬁle in the law of the wall: a positive $\Delta U^+$ corresponds to drag reduction, and a negative $\Delta U^+$ implies drag increase, as for the conventional $k$-type roughness. Part of the trends seen in figure \ref{fig:Re} for drag reduction data are due to Reynolds effects; properly removing them via analytical relations is possible, as done in ref.\cite{gatti-quadrio-2016} for active spanwise forcing, and would contribute to clarifying the situation, by exposing some remaining "puzzling" trends with $Re$ (to cite words used in ref.\cite{spalart-etal-2019}).

\section{Conclusions}
\label{sec:conclusions}

In this review paper we have provided a brief and up-to-date description of what we know and what we don't about the potential of dimples for turbulent skin-friction drag reduction. 
While we can't obviously offer an answer to the still-standing question whether or not dimples are a suitable technique to reduce turbulent skin-friction drag, it is our hope that this comprehensive overview will at least help the newcomer to frame the problem, quickly identify the key references, and get a glimpse at the complexity of the topic. 

While reviewing the state of the art, we have also mentioned some methodological issues that bear a critical importance when attempting to measure drag changes by dimples. Leveraging concepts and procedures (and perhaps facilities altogether) developed over the years for riblets might yield data that are reliable enough to begin understanding the physics behind dimple drag reduction, a necessary and preliminary step to improve their performance.

\appendix
\section[\appendixname~\thesection]{Computational details}
\label{sec:appendix}

In this review we have also presented results from DNS simulations carried out on purpose for the present work. They concern a turbulent plane channel flow, with dimples placed on one wall only. The employed parallel DNS code was introduced by \cite{luchini-2016}, and solves the incompressible Navier--Stokes equations in primitive variables on a staggered Cartesian grid. Space discretization is based on second-order finite differences, and temporal integration uses a fractional time stepping method based on a third-order Runge--Kutta scheme. The Poisson equation for the pressure is solved by an iterative successive over-relaxation algorithm. An implicit immersed-boundary method, implemented in staggered variables, continuous with respect to boundary crossing and numerically stable at all distances from the boundary \cite{luchini-2013, luchini-2016}, describes the geometry of the non-planar wall. 
Periodic boundary conditions are enforced in both the streamwise and spanwise directions, while no slip and no penetration boundary conditions are enforced at the walls.

The size of the computational domain (and therefore the number of dimples considered) is chosen to ensure that it is always larger than the minimal flow units needed to sustain the near-wall turbulence cycle \cite{jimenez-moin-1991}. The smallest domain in our simulations has size $L_x=4\sqrt{3}h$ and $L_z=4h$ in external units and $L_x^+=1385$ and $L_z^+=800$ in viscous units. A uniform distribution of points is used in both the streamwise and spanwise directions, with the selected grid spacing ensuring that $\delta x^+ \lessapprox 10$ and $ \delta z^+ \lessapprox 5$ for all the considered cases. In the wall-normal direction a non-uniform distribution is used to properly resolve the dimples and the near wall region. The grid spacing is indeed constant from the dimple bottom to the dimple tip, from where a hyperbolic tangent distribution is used. The number of points in the wall-normal direction is chosen to ensure that at the walls $\delta y^+ < 1$ for all cases. The number of points for the simulations in figure \ref{fig:origin}, carried out at about $Re_b=2800$ (or $Re_\tau=180$) is $N_x=260$, $N_y=260$ and $N_z=260$. For the simulations in figure \ref{fig:deep-dimples}, instead, the number of points increases up to $ N_x=300$, $N_y=334$ and $N_z=300$ to deal with the higher Reynolds numbers, since in this dataset the Reynolds number varies, in the range $2690 \le Re_b \le 10450$ (or $160 \le Re_\tau \le 550$).


\bibliographystyle{plain}
\bibliography{../../Wallturb.bib}

\end{document}